\documentclass[10pt,showpacs,preprintnumbers,amsmath,amssymb,
aps,prd,nofootinbib,eqsecnum,a4paper]{revtex4}

\usepackage{graphicx,epsf}

\newcommand{\be}{\begin{equation}}
\newcommand{\ee}{\end{equation}}
\newcommand{\bea}{\begin{eqnarray}}
\newcommand{\eea}{\end{eqnarray}}

\renewcommand{\geq}{\geqslant}
\newcommand{\p}{\partial}
\newcommand{\n}{\nabla}

\newcommand{\R}{\mathcal{R}}
\renewcommand{\H}{\mathcal{H}}
\newcommand{\K}{\mathcal{K}}
\newcommand{\vp}{\varphi}
\newcommand{\dvp}[1]{{\delta\varphi_{#1}}}

\newcommand{\eq}[1]{Eq.~(\ref{#1})}

\newcommand{\Planck}{M_{\mathrm{P}}}
\renewcommand{\d}{\mathrm{d}}
\newcommand{\fnl}{f_{\mathrm{NL}}}
\newcommand{\tnl}{\tau_{\mathrm{NL}}}
\DeclareMathOperator{\tr}{tr}

\begin{document}
\preprint{}
\title{Different approaches to the second order Klein--Gordon equation}
\author{Karim A.~Malik}
\affiliation{Astronomy Unit, School of Mathematical Sciences,
Queen Mary, University of London, Mile End Road, London E1 4NS,
United Kingdom}
\author{David Seery}
\affiliation{Centre for Theoretical Cosmology,
Department of Applied Mathematics and Theoretical Physics,
University of Cambridge, Wilberforce Road, Cambridge, CB3 0WA,
United Kingdom}
\author{Kishore N.~Ananda}
\affiliation{Cosmology and Gravity group, Department of Mathematics
and Applied Mathematics, University of Cape Town, Rondebosch 7701,
Cape Town, South Africa}
\pacs{98.80.Cq}
\date{\today}

\begin{abstract}
We derive the Klein--Gordon equation for a single scalar field coupled
to gravity at second order in perturbation theory and leading order in
slow-roll. This is done in two ways: we derive the Klein--Gordon
equation first using the Einstein field equations, and then directly
from the action after integrating out the constraint equations.
We also point out an unexpected result regarding the treatment of the
field equations.
\end{abstract}

\maketitle

\section{Introduction}

Cosmological perturbation theory is an essential tool which connects
theories of the early universe with observation. Until recently
it was sufficient to use linear perturbation theory to make and compare
theoretical predictions with Cosmic Microwave Background (CMB)
experiments and data from large scale structure surveys. One of the
major breakthroughs at first order was the development of the
gauge-invariant formalism, initiated by Bardeen and others
\cite{Bardeen80,KS,MFB,LLBook}.
The next round of CMB observations, which will be carried by
observatories such as \emph{Planck}, together with data from
anisotropies in matter fields such as the neutral hydrogen, will
require gauge-invariant \emph{second order} perturbation theory. A
vigorous effort is therefore under way to develop higher-order
perturbation theory, and change it from what was once of mere academic
interest and a calculational nuisance, into a powerful tool, see for
example Refs.~\cite{Mukhanov,Gauge1,Met2,Noh,MW2004,Rigopoulos:2004gr,
Bartolo:2004if,Bart1,Nak3,LMS,Tom1,Lyth:2005du,
Rigopoulos:2005xx,Malik2005,Bart2,Nak5,Malik2006,Cov1,
Ananda1,Mena1,Ananda2}.

Considerable progress has already been made. Within the last five
years, reliable predictions have become available for the degree of
non-linearity imprinted by inflation in the CMB temperature anisotropy
beyond linear order,
e.g.~\cite{Maldacena,BartNG5,Seery:2005wm,Seery:2005gb,Vernizzi:2006ve,
Seery:2006vu,Battefeld:2006sz,Seery:2006js,Battefeld:2007en}.
Indeed, there is already some interesting tension between large
classes of theoretically well-motivated models and observation
on the basis of the non-linearity they predict
\cite{Lidsey:2007gq,Baumann:2006cd,Peiris:2007gz,Bean:2007eh}.
For example, one of the strongest bounds on the parameter space in the
curvaton scenario comes from the limit on the non-gaussianity
parameter $f_{\rm{NL}}$
\cite{Lyth:2002my,Bartolo:2003jx,Malik:2006pm,Sasaki:2006kq}. However,
despite the considerable effort expended on the development of the
theory, some interesting surprises can be found in unexpected
``corners''.

Much of this research has been framed in the context of the Lagrangian
formalism, where one begins with an \emph{action} and quantises it. In
doing so, one encounters a variant of the Feynman diagrams which are
used to compute scattering amplitudes in particle physics. This method
of calculation has the advantage that a good deal of pre-existing
knowledge can be imported wholesale from high-energy physics.
Its principal drawback arises when making comparisons with the
cosmological literature, which has traditionally approached perturbation
theory from the standpoint of the Einstein equations.

In this paper we explore the relationship between these two approaches
by returning to the Einstein equations and calculating the Klein--Gordon
equation at second order in perturbation
theory, specialising to the case of a single scalar field and working
to first order in the slow-roll approximation.
This result has already been published in
Ref.~\cite{Malik2006} in the general case, but here we relate it to a
different method of derivation, namely variation of the Einstein action
coupled to a scalar field.
In doing so we
highlight a potential pitfall in the reduction of the full Einstein
equations, which if unobserved can lead to errors at second order. In
particular, this means that techniques which have been applied
successfully at first order \cite{MFB} may need treating with
more care once one incorporates the effect of non-linearities.
We obtain consistent results in both cases.

Throughout this paper we use natural units where
$8\pi G\equiv \Planck^{-1/2}$ is set equal to unity, where $\Planck$
is the so-called reduced Planck mass.
Derivatives with respect to conformal time are denoted by a
prime. Greek indices, $\mu,\nu,\lambda$, label spacetime coordinates and
run from $0$ to $3$, while lower case Latin indices, $i,j,k$,
label purely spatial coordinates and run from $1$ to $3$.
The metric convention is $(-,+,+,+)$ and the background spacetime
is taken to be of Friedmann--Robertson--Walker (FRW) form,
\begin{equation}
    \d s^2 = g_{\mu\nu} \d x^\mu \d x^\nu
        = - \d t^2 + a^2 \delta_{ij} \d x^i \d x^j \,,
\end{equation}
where $a = a(t)$ is the scale factor. It is usually more convenient to
work in so-called conformal time, defined by
$\eta = \int_{t}^{\infty} \d t / a(t)$.
Note that throughout this paper we work in a spatially flat background
spacetime, and evaluate all perturbed quantities on uniform curvature
hypersurfaces~\cite{Malik2005}.

The paper is organised as follows. In the following section we define
the metric and and the matter content of our spacetime. In Section
\ref{KGusingfield} we derive the Klein--Gordon equation using the field
equations and highlight possible difficulties involving the
spatial part of the second order field equations. In Section \ref{KGusingaction}
we derive the Klein--Gordon equation directly from the action.
We conclude with a discussion in Section~\ref{discussion}.

\section{Preliminaries}

In this section we define the metric and the matter variables we will
use in the following sections.

\subsection{Notation}

We begin by defining the metric tensor in the uniform curvature gauge
up to second order in the perturbations, including scalar
perturbations only,
\bea
\label{metric1}
g_{00} & = & -a^2\left(1+2\phi_1+\phi_2\right) \,, \qquad
g_{0i} = a^2\left(B_1+\frac{1}{2}B_2\right)_{,i}\,, \qquad
g_{ij} = a^2\delta_{ij}\,,
\eea
where
$\phi_1$ and $\phi_2$ together describe the
lapse function, and $B_1$ and $B_2$ comprise the scalar part of the shift
function (describing the the shear in this gauge). Numerical
subscripts denote the
order of the perturbation, and a comma denotes a partial derivative,
that is, $X_{,i} \equiv \p X/\p x^i$.
The notation used in the present paper coincides with the conventions
used, for example, in Ref.~\cite{Malik2006}. In this notation,
any tensorial quantity is expanded in a truncated power series, as shown above
for the metric tensor (\ref{metric1}).
Similarly the scalar field $\vp$ can be split into a
background part and a perturbation, which we take to be
\be
\vp\equiv \vp_0+\dvp1+\frac{1}{2}\dvp2\,,
\ee
where $\vp_0$ is the homogeneous background, $\dvp1$ is chosen to obey
Gaussian statistics, and the higher terms $\dvp{n}$, $\phi_{n}$ and
$B_{n}$ for $n \geq 2$ are polynomials in $\dvp1$ of order $n$.
Note that since we work in the flat gauge, the field fluctuation at
first and second order is merely the
associated Sasaki--Mukhanov variable, \cite{Sasaki1986,Mukhanov88,MW2004}.
The authors of Refs.~\cite{Seery:2005wm,Seery:2005gb} use a different
convention for the perturbations. However, by observing that their
field perturbation $\delta\phi_{\rm{SL}}$ is related to our
perturbations $\dvp1$ and $\dvp2$ by
$\delta\phi_{\rm{SL}}\equiv\dvp1+\frac{1}{2}\dvp2$, we can readily
compare our results.

\subsection{Action and energy--momentum tensor}

The action for a single scalar field minimally coupled to standard
Einstein gravity is \cite{Boundary,Chamblin:1999ya}
\be
    \label{action}
    S=\int \d^4 x\sqrt{-g}\left(\frac{1}{2}\R+{\cal{L}}_\vp\right)
    + \int_{\partial} \d^3 x \sqrt{h} \; \K\,,
\ee
where $g_{\mu\nu}$ is the spacetime metric, $g \equiv \det g_{\mu\nu}$
is its determinant, $\R$ is the spacetime Ricci scalar, and $\partial$
denotes the boundary of spacetime, if one exists. If a boundary is
present, then $h_{ij}$ is its first fundamental form (also referred to
as the induced metric, obeying $h_{ij}\equiv g_{ij}$ in the case of
the boundary being a hypersurface of constant time), and $\K_{ij}$ its
second fundamental form (or extrinsic curvature), where
$\K = \tr\K_{ij}$.

The matter Lagrangian is given by
\be
\label{Lscal}
{\cal{L}}_\vp =
-\frac{1}{2}g^{\mu\nu}\varphi_{,\nu}\varphi_{,\mu}
-U(\varphi) \,,
\ee
where $\vp$ is the scalar field, $U$ is its potential,
and $\vp_{,\mu}\equiv \p\vp / \p x^\mu$.
The energy--momentum tensor is obtained by varying the action with respect
to the metric, giving
\be
T_{\mu\nu}=-2\frac{\partial {\cal{L}}_\vp}{\partial g^{\mu\nu}}
+g_{\mu\nu} {\cal{L}}_\vp \,.
\ee
When the field equations are satisfied, this is automatically
conserved as a consequence of Noether's theorem.  For a single scalar
field, $T_{\mu\nu}$ takes the form
\be
\label{multiTmunu}
T_{\mu\nu}
=\left[\vp_{,\mu}\vp_{,\nu}
-\frac{1}{2}g_{\mu\nu}g^{\alpha\beta}\vp_{,\alpha}\vp_{,\beta}
\right] - g_{\mu\nu} U(\vp) \,.
\ee
%

\section{Deriving the second order Klein--Gordon equation
using the field equations}
\label{KGusingfield}

We now proceed to derive the governing equation for the scalar field,
namely the Klein--Gordon equation, using the field equations. We
therefore start by writing down Einstein's field equations
\be
    \label{einstein}
    G_{\mu\nu}=T_{\mu\nu}\,,
\ee
which can be found by varying the action (\ref{action}). For this purpose
the inclusion of the Gibbons--Hawking term on the boundary is essential.
The Bianchi identities then imply the conservation of energy--momentum,
\be
    \label{energy}
    \nabla^\mu T_{\mu\nu}=0\,,
\ee
where $\nabla^\mu$ denotes the covariant derivative.

\subsection{Field dynamics}
\label{field_dyn}

The energy conservation equation, Eq.~(\ref{energy}),
gives an evolution equation for the $\dvp{n}$ order-by-order (see
e.g. Ref.~\cite{Malik2005} for details).
At zeroth order in perturbations, one finds the background equation
\bea
\label{KGback}
\vp_{0}''+2\H\vp_{0}'+a^2 U_{,\vp}=0\,,
\eea
where $\H\equiv a'/a$ and is related to the Hubble parameter
$H$ by $\H=aH$. At first order we find
\be
\label{KG1_raw}
\dvp1''+2\H\dvp1'-\nabla^2\dvp1+2a^2 U_{,\vp}\phi_1
-\vp_{0}'\nabla^2 B_1
-\vp_{0}'\phi'_1+a^2 U_{,\vp\vp}\dvp1
=0\,,
\ee
whereas after some calculation the second order equation becomes
\bea
\label{KG2_raw}
\dvp2''&+&2\H\dvp2'-\nabla^2\dvp2+a^2 U_{,\vp\vp}\dvp2
+a^2  U_{,\vp\vp\vp} \dvp1\dvp1 +2a^2 U_{,\vp}\phi_2
-\vp_{0}'\left(\nabla^2 B_2+\phi_2'\right)\nonumber\\
&+&4\vp_{0}' \p_kB_{1}\p^k\phi_{1} +2\left(2\H\vp_{0}'+a^2
U_{,\vp}\right) \p_kB_{1}\p^k B_{1} +4\phi_1\left(a^2
U_{,\vp\vp}\dvp1-\nabla^2\dvp1\right)
+4\vp_{0}'\phi_1\phi_1'\nonumber\\ 
&-&2\dvp1'\left(\nabla^2 B_1+\phi_1'\right)-4\p_k\dvp1'\p^kB_{1} 
=0\,.  
\eea

In order to arrive at an equation which can reasonably be thought of as a
second-order version of the Klein--Gordon equation, we must eliminate the
terms $\{ B_1, B_2, \phi_1, \phi_2 \}$ which come from the metric
and rewrite them in terms of the field fluctuations $\{ \dvp1, \dvp2 \}$.
This is done by using the constraint part of the Einstein equations.
Since we will use the slow-roll approximation to control this part of the
calculation, it is first necessary to decide to what order in the
approximation our calculations must be carried.

The slow-roll expansion can be thought of as an expansion in powers of
$2 \epsilon \equiv (\vp_0'/\H)^2$, together with other small
quantities obtained by differentiating $\epsilon$ with respect to
cosmic time $t$.  It is easy to see from Eq.~(\ref{KG2_raw}) that all
intrinsically second order terms (namely, $B_2$ and $\phi_2$) are
accompanied by a factor of $\vp_0'$, whereas some terms formed from
the product of first order quantities appear without any factors of
the time derivative of the background field.
Therefore we will aim to compute only to $\vp_0'/\H \sim \sqrt{\epsilon}$,
for which purpose it is sufficient
to obtain second order quantities only to zeroth order in slow-roll,
but we must carry the calculation of first order quantities to the
first non-trivial order.

When $\epsilon \ll 1$, one says that the field is in the slow-roll
regime. Extracting the leading slow-roll part of the background equation,
\eq{KGback}, we obtain
\be \label{KGslowroll}
3\H\vp_{0}'+a^2 U_{,\vp}=0\,.
\ee

\subsection{Einstein equations}
\label{Einstein_sec}

We now turn to the Einstein equations. These break into ten independent
equations, of which some are evolution equations and some are constraints.
In the present case there is only one degree of freedom, the field
fluctuation $\{ \dvp1, \dvp2 \}$, and therefore there will ultimately
be only one non-trivial evolution equation.
In this section we aim to use the remaining equations, which are constraints,
to obtain the metric quantities $\{ B_1, B_2, \phi_1, \phi_2 \}$
in terms of the gaussian field fluctuation, $\dvp1$.

\subsubsection{Background}

At zeroth order in slow roll the background Einstein equations
imply the Friedmann equation, which simplifies to
\be
\label{Friedmann}
\H^2=\frac{1}{3} a^2 U_0\, .
\ee
This gives an evolution equation for $\H$, namely $\H'=\H^2$.
%

\subsubsection{First order}
\label{first_order_field}

Now consider the terms which are first order in perturbations.
The $(0,0)$ component of the Einstein equations is
\be
\label{00Ein1v1}
2a^2U_0{\phi_1}+ \vp_{0}'{{\dvp1}}'+a^2{\delta U_1}
+2{\H}\nabla^2B_1
=0\,,
\ee
and the $(0,i)$ part gives
\be 
\label{0iEin1} 
\p_i\left(2\H\phi_1-\vp_{0}'{{\dvp1}}\right)=0\,,
\ee
where the gradient can readily be removed, e.g.~by working Fourier space.
Working to first order in slow roll and using the background equations
given above, we find that \eq{00Ein1v1} can be rewritten as
\be
\label{00Ein1v2}
\vp_{0}'{{\dvp1}}'+2{\H}\nabla^2B_1
=0\,.
\ee
This is a constraint which allows us to eliminate $B_1$ from
Eq.~(\ref{KG2_raw}) in favour of of the field fluctuation $\dvp1$.

The $(i,j)$ component of the Einstein equation is given by
\be
\label{Einij_full1}
\left(2\H\phi_1'-\vp_0'\dvp1+2a^2 U_0\phi_1+a^2\delta U_1\right)\delta_{ij}
+\left(\nabla^2\delta_{ij}-\p_i\p_j\right)\left(B_1'+2\H B_1+\phi_1\right)
=0\,.
\ee
Taking the trace of \eq{Einij_full1} gives one scalar equation,
\be
\label{trace1raw}
3\left(2\H\phi_1'-\vp_0'\dvp1+2a^2 U_0\phi_1+a^2\delta U_1\right)
+2\nabla^2\left(B_1'+2\H B_1+\phi_1\right)=0\,.
\ee
At linear
order we have now two options to ``extract'' another scalar equations
from \eq{Einij_full1}:
\begin{itemize}
\item
It is possible to simply read off the off-diagonal (or $i \neq j$)
part of the $(i,j)$ equation, \eq{Einij_full1}. This method was used
by Mukhanov, Feldman \& Brandenberger \cite{MFB} at first order, in
conjunction with invariance arguments which allowed them to extend the
three independent $i \neq j$ equations to a single scalar equation
which constrained the $i \neq j$ term to vanish.
\item
Another possibility, advocated here, is to apply the operator $\p^i
\p^j$ to \eq{Einij_full1}. The benefit in doing so is that one arrives
at a scalar equation via manipulations which are manifestly
tensorial. Observe that the operator $\nabla^2\delta_{ij}-\p_i\p_j$ is
transverse to $\p^i \p^j$ and therefore vanishes, leaving an evolution
equation for $\phi_1$.
\end{itemize}
Although both methods yield valid results at first order, it is not
clear how the former should be extended
to second order. On the other hand, the
second method requires only textbook manipulations, leading to a
correct scalar equation via an essentially mechanical process. Its
principal disadvantage is the necessity to introduce inverse Laplacian
operators.
In either case, one can extract an evolution equation for the shear,
\be
\label{shear1}
\n^2\left(B_1'+2\H B_1+\phi_1\right)=0\,,
\ee
and an evolution equation for the lapse function
\be
\label{trace1refined}
2\H\phi_1'+
a^2\delta U_1+2a^2U_0{\phi_1}- \vp_{0}'{{\dvp1}}'=0\, .
\ee
After dropping terms which are higher order in slow-roll this
simplifies substantially, yielding
\be
\label{trace1}
2\H\phi_1'- \vp_{0}'{{\dvp1}}'=0\,,
\ee
Any combination of Eq.~(\ref{00Ein1v2}) with Eqs.~(\ref{shear1})
or~(\ref{trace1}) allows us to determine $\phi_1$ in terms of
$\dvp1$. We therefore have both metric potentials which are necessary
to rewrite Eq.~(\ref{KG1_raw}) as a single equation for $\dvp1$.

\subsubsection{Second order}

As pointed out in Section \ref{field_dyn} above we only require the
second order field equations to zeroth order in slow roll.  Reducing
the calculation to the leading order slow-roll effect considerably
simplifies both the calculation and presentation of our results.

The $(0,i)$ Einstein equation is given by
\be
\label{0i_2}
\p_i\H\phi_{2}-\dvp1'\p_i\dvp1=0
\ee
This can be rewritten as a scalar equation by taking its divergence,
yielding
\bea
\label{0i_2version2}
&&\H\phi_{2}
- \nabla^{-2}\left(
\dvp1'\nabla^{2}\dvp1+\p_k\delta\vp_{1}'\p^k\delta\vp_{1}
\right)=0\,,
\eea
where we introduce the inverse Laplacian,
which satisfies the identity $\nabla^{-2}(\nabla^{2})X=X$.

The second order $(i,j)$-equation is given by
\be
\label{Einij_full2}
\left(2\H\phi_2'+2a^2U_0\phi_2\right)\delta_{ij}
+\left(\nabla^2\delta_{ij}-\p_i\p_j\right)\left(B_2'+2\H B_2+\phi_2\right)
+\left(\p_k\dvp1\p^k\dvp1-{\dvp1'}^2\right)\delta_{ij}
-2\p_i\dvp1\p_j\dvp1=0\,.
\ee
The spatial trace is obtained from \eq{Einij_full2} and satisfies
\bea
\label{spatialtrace2}
&&3\H\phi_2'+3 a^2 U_0 \phi_2
+\nabla^2\left(B_2'+2\H B_2+\phi_2\right)
+\frac{1}{2}\left(
-3{\dvp1'}^2
+\p_k\dvp1\p^k\dvp1\right)=0\,.
\eea

Following the discussion above, we now let $\p_i\p_j$ act on
\eq{Einij_full2} to extract another scalar equation from the $(i,j)$
part, and observe that
$\p^i\p^j\left(\nabla^2\delta_{ij}-\p_i\p_j\right)X=0$. This gives
\be
\label{phi2dash}
\H\phi_2'+3\H\nabla^{-2}\left(\dvp1'\nabla^{2}\dvp1
+\p_k\dvp1'\p^k\dvp1\right)
-\nabla^{-2}\left(\nabla^{2}\dvp1\nabla^{2}\dvp1
+\dvp1'\nabla^{2}\dvp1'+\p_k\dvp1\nabla^{2}\p^k\dvp1
+\p_k\dvp1'\p^k\dvp1^{\prime}\right)=0\,.
\ee

Note that this usefulness of this procedure
is not specific to our choice of gauge or
dependent on invoking the slow roll approximation.
We believe it is simplest to employ our method
to extract a scalar equation from the
$(i,j)$ part of the Einstein equation, (\ref{Einij_full2}), whenever
the equation can not be written as a total derivative. This is in
general the case at second order and higher, when terms of the form
$X_{1i}Y_{1i}$ etc. will appear.

How can we be confident
that we are proposing the correct method? The first
reason is that only this approach leads to a consistent set of
Einstein equations.
%
The other is that we get the same result for the Klein--Gordon equation
using the action method, as shown in Section
\ref{KGusingaction}.

Since equation (\ref{phi2dash}) is just an evolution equation for the
second order lapse function, another way of deriving it, is to take
the time derivative of the constraint (\ref{0i_2version2}).
In this case we also need the first order Klein--Gordon equation in the
extreme slow roll limit, which follows from \eq{KG1_raw} as
\be
\label{KG1extreme}
\dvp1''+2\H\dvp1'-\nabla^2\dvp1=0\,.
\ee
Substituting \eq{KG1extreme} into the time derivative of
\eq{0i_2version2}, we recover \eq{phi2dash}, as required.

\subsection{The Klein--Gordon equation at second order}

We can now substitute the field equations at first and second order
into the Klein--Gordon equation (\ref{KG2_raw}). Working to leading order
in slow roll we get
\bea
\label{KG2_field}
\dvp2''&+&2\H\dvp2'-\nabla^2\dvp2 +a^2  U_{,\vp\vp\vp} \dvp1^2 \nonumber\\
&+&\frac{\vp_{0}'}{\H}\Bigg\{
\frac{1}{2}\left({\dvp1'}^2+\p_k\dvp1\p^k\dvp1\right)
-2\dvp1\nabla^2\dvp1
-\nabla^{-2}\Big[\nabla^2\dvp1\nabla^2\dvp1+\dvp1'\nabla^2\dvp1'
\nonumber\\
&&\hspace{44mm}
+\p_k\dvp1\nabla^2\p^k\dvp1+\p_k\dvp1'\p^k\dvp1^{\prime}
\Big]
+2\p_k\dvp1'\nabla^{-2}\p^k\dvp1^{\prime}\Bigg\}
=0\,.
\eea
When applying the slow-roll expansion, one ordinarily thinks of the
higher derivatives of the potential, such as
$U_{,\vp\vp\vp}$, as being negligible in comparison with the leading
slow-roll terms. In Eq.~(\ref{KG2_field}) this term has been
retained.  This is because although $U_{,\vp}/U \ll 1$ is necessary in
order for inflation to occur at all, and $U_{,\vp\vp}/U \ll 1$ is
necessary in order to have sufficient e-foldings, there
are no such constraints for higher derivatives.  Therefore it may well
be that models exist in which $U_{,\vp\vp\vp}$ is unusually
large. In such a model, $U_{,\vp\vp\vp}$ will make a contribution to
the non-gaussianity imprinted in the cosmic microwave background which
cannot be ignored \cite{Zaldarriaga}.

\section{Derivation from the third order action}
\label{KGusingaction}

Let us now return to the action and derive the Klein--Gordon equation
from it. This follows a somewhat different procedure than that used
with the field equations, owing to the way constraints are implemented
in the action formalism.

The action is found to depend only algebraically on the lapse and shift
functions and their spatial derivatives; it involves no time derivatives
of these quantities. Since one finds the equation of motion for each
field, say $W$, by varying the action with respect to $W$ and demanding
that the resulting variation $\delta S/\delta W$ is zero, the equations
of motion for the lapse $\phi$ and the shift $B$ are also purely algebraic.
They can be solved to give expressions in terms of the field fluctuation
which hold at all times and do not require integrating any equation of motion.
Once $\phi$ and $B$ have been determined, they can be eliminated from
the Lagrangian. Varying the resulting action with respect to the field
gives the equation of motion directly, which must coincide with the
Klein--Gordon equation derived above.

This calculation was initially given (in the flat and comoving
slicings) by Maldacena \cite{Maldacena} and refined in the flat
slicing in Ref.~\cite{Seery:2005gb}. In conformal time, the part of
the action quadratic in $\dvp2$ can be written
\begin{equation}
    S_2 = \frac{1}{8} \int \d\eta \, \d^3 x \; a^2 \Big\{
        (\dvp2')^2 - (\partial\dvp2)^2 \Big\} .
    \label{action-quadratic}
\end{equation}
On the other hand, there is also an ``interaction'' term which
involves a product of the $\dvp{n}$. The leading term in this interaction
is linear in $\dvp2$ and quadratic in $\dvp1$,
\begin{equation}
    S_3 = \int \d\eta \, \d^3 x \; a^2 \left[
        \frac{1}{3!} U_{,\vp\vp\vp} \dvp2 (\dvp1)^2 +
        \frac{\vp_0'}{4\H}\left\{
            \dvp2' \p^k \nabla^{-2} \dvp1' \p_k \dvp1 -
            \frac{1}{2} \dvp2 \left[ (\dvp1')^2 + (\partial \dvp1)^2 \right]
        \right\}
    \right]+ \mbox{permutations} ,
    \label{action-cubic}
\end{equation}
where the permutations are formed by swapping the position of $\dvp2$ among
the three possible locations.
The higher derivative $U_{,\vp\vp\vp}$ has been
included to account for the possibility that it is
unusually large, as discussed above.
The field equation for $\dvp2$ is obtained
by demanding that $\delta S/\delta(\dvp2) = 0$, ignoring any boundary terms.

Consider first $S_2$. It is clear that $\delta S_2/\delta(\dvp2)$
can be written
\begin{equation}
	\label{quad-variation}
    \delta S_2 =
    \frac{1}{4} \int \d \eta \, \d^3 x \; \delta(\dvp2) \;
    a^2 ( - \dvp2'' - 2\H \dvp2' + \nabla^2 \dvp2 )
    + \frac{1}{4} \int_\partial \d^3 x \; a^2 \delta(\dvp2) \dvp2' .
\end{equation}
Now consider $S_3$. From left to right this breaks into four terms,
the first of which is the variation of the $U_{,\vp\vp\vp}$ term and
is trivial.
The variation of the second term gives (including the effect of permuting
the location of $\dvp2$)
\begin{eqnarray}
    \nonumber
    \delta S_3 & \supseteq &
    \int_{\partial} \d^3 x \; a^2 \frac{\vp_0'}{4\H} \delta(\dvp2) \left\{
    	\partial^k \nabla^{-2} \dvp1' \partial_k \dvp1 -
    	\nabla^{-2}\left[
    		\partial^k \dvp1' \partial_k \dvp1 +
    		\dvp1' \nabla^2 \dvp1
    	\right] \right\}
    \\ & & \nonumber \mbox{} +
    \int \d \eta \, \d^3 x \; a^2 \frac{\vp_0'}{4\H} \delta(\dvp2) \left\{
        \frac{4}{a^2} \p^k \dvp1 \p_k \nabla^{-2} \frac{\delta S_2}{\delta(\dvp1)} -
        \frac{4}{a^2} \nabla^{-2} \left[ (\nabla^2 \dvp1 + \p^k \dvp1 \p_k)
            \frac{\delta S_2}{(\dvp1)}
        \right]
    \right\}
    \\ & & \nonumber \mbox{} +
    \int \d \eta \, \d^3 x \; a^2 \frac{\vp_0'}{4\H} \delta(\dvp2) \Bigg\{
        \nabla^{-2} \left[
            \nabla^2 \dvp1 \nabla^2 \dvp1 + \p^k \dvp1 \p_k \nabla^2 \dvp1 +
            \p^k \dvp1' \p_k \dvp1' + \dvp1' \nabla^2 \dvp1' \right]
            \\ & & \hspace{40mm} \mbox{} -
            \p^k \dvp1 \p_k \dvp1 - 2 \p^k \dvp1' \p_k \nabla^{-2} \dvp1'
            - (\dvp1')^2 \Bigg\} .
	\label{cubic-variation}
\end{eqnarray}
The terms proportional to $\delta S_2/\delta(\dvp1)$ vanish when they
are evaluated on the classical solution, which is obtained by demanding
that Eq.~(\ref{quad-variation}) is zero. For this purpose one ignores
the surface term, choosing boundary conditions for $\delta(\dvp1)$ such that
it vanishes. If one chooses to retain the surface term then the
corresponding surface terms in Eq.~(\ref{cubic-variation}) are eliminated,
but any gain is illusory since they must then be added back in
from the equation of motion for $\dvp1$.

The third term in \eq{action-cubic} gives a variation
\begin{equation}
    \delta S_3 \supseteq
    \int_{\partial} \d^3 x \; a^2 \frac{\vp_0'}{4\H} \delta(\dvp2) \left\{
    	- \dvp1 \dvp1' \right\} + 
    \int \d \eta \, \d^3 x \; a^2 \frac{\vp_0'}{4\H} \delta(\dvp2) \left\{
        - \frac{4}{a^2} \dvp1 \frac{\delta S_2}{\delta(\dvp1)} +
        \dvp1 \nabla^2 \dvp1 + \frac{1}{2} (\dvp1')^2 \right\} ;
\end{equation}
and, likewise, the fourth term contributes
\begin{equation}
    \delta S_3 \supseteq \int \d \eta \, \d^3 x \; a^2
        \frac{\vp_0'}{4\H} \delta(\dvp2) \left\{
        \frac{1}{2} \p^k \dvp1 \p_k \dvp1 + \dvp1 \nabla^2 \dvp1
    \right\} .
\end{equation}
These terms all combine to give the total variation. Discarding
those terms proportional to $\delta S_2/\delta(\dvp1)$
together with all surface terms,
one finds the overall Klein--Gordon equation
\begin{eqnarray}
\label{KG2_action}
    \nonumber
    \dvp2'' & + & 2\H \dvp2' - \nabla^2 \dvp2 + U_{,\vp\vp\vp} (\dvp1)^2 =
    \\ & + & \nonumber
    \frac{\vp_0'}{\H} \Big\{
        - \frac{1}{2} \p^k \dvp1 \p_k \dvp1 - \frac{1}{2} (\dvp1')^2 +
        2 \dvp1 \nabla^2 \dvp1 - 2 \p^k \dvp1' \p_k \nabla^{-2} \dvp1'
    \\ & & \hspace{20mm} \mbox{} +
        \nabla^{-2} \left[
            \nabla^2 \dvp1 \nabla^2 \dvp1 + \p^k \dvp1 \p_k \nabla^2 \dvp1 +
            \p^k \dvp1' \p_k \dvp1' + \dvp1' \nabla^2 \dvp1' \right]
    \Big\} .
\end{eqnarray}
Equation (\ref{KG2_action}) agrees exactly with the Klein--Gordon equation
which was found using the Einstein equations, \eq{KG2_field}.

\section{Discussion}
\label{discussion}

In this paper, we have derived the Klein--Gordon equation at second order
in the perturbations and leading order in slow-roll. We have shown
explicitly for the first time that the two most popular approaches in the
literature, based on the Einstein field equations and the action
principle, yield equivalent results. This goes some way to demonstrating
the consistency of current beyond-leading-order calculations,
including those of the non-linearity parameters $\fnl$ and $\tnl$
\cite{Maldacena,Seery:2005gb,Seery:2006vu}
which will be important for future CMB experiments.

In many cases one can choose freely whether the action principle or the
Einstein field equations are more appropriate for the task in hand.
In the case of the Einstein equations, one needs the relevant constraint
equations at $n$th order, and an evolution equation. In principle this
is contained in the Einstein equations themselves, but in practise
it is often more convenient to obtain an evolution equation from the
Bianchi identities, as we have done in the present paper. For the
action, one must compute the Lagrangian to $(n+1)$th order, and then
eliminate the constraints associated with the lapse function and shift
vector. However, owing to the precise way these
appear in the action, there is a special simplification which implies
one need obtain explicit expressions only to $(n-1)$th order.
The result, at least for low orders of
perturbation theory, is that the amount of work required in both
approaches seems much the same.

These two methods are traditionally used for different purposes.
The field equations are ordinarily the method of choice for obtaining
numerical solutions. On the other hand, the action is typically preferred
when one aims to compute scattering amplitudes, or more generally the
correlation functions of a quantised theory. This is because the action,
being a scalar, manifestly exhibits the full group of symmetries
associated with the theory. Moreover, it facilitates the use of the
so-called ``interaction picture'' which is the most common approach for
dealing with fields beyond first order in perturbation theory.

In some cases, however, there is no choice and one must work with the
equations of motion. For example, this is the case with dissipative systems
such as fluids for which it is difficult or impossible to formulate an
action principle. In this case one would like to extend the existing
second-order framework to allow the computation of quantities
traditionally associated with the action, such as the correlation functions
of the quantised theory. The analysis given in this paper is a first step
in that direction.

We have pointed out that some care is required in reducing the full set
of Einstein equations. In general this necessitates contracting spatial
tensors into scalar equations by the application of a suitable derivative
operator. Together with the spatial trace, this gives two scalar
equations which at first order coincide with the well-known standard
results \cite{MFB}. At second order, the analysis requires some more
care in order to obtain correct results.

\acknowledgments

The authors are grateful to Jim Lidsey, David Lyth and David
Matravers for useful discussions and comments.
KNA and DS are grateful to the Astronomy Unit at QMUL for their
hospitality while visiting.
KNA is supported by the National Research Foundation (South Africa)
and the {\it Ministrero deli Affari Esteri- DIG per la Promozione e
Cooperazione Culturale} (Italy) under the joint Italy/South Africa
science and technology agreement.
%

{}


\begin{thebibliography}{}

\bibitem{Bardeen80}
J.~M.~Bardeen,
Phys.\ Rev.\ D {\bf 22}, 1882 (1980).

\bibitem{KS}
H.~Kodama and M.~Sasaki,
``Cosmological Perturbation Theory,''
Prog.\ Theor.\ Phys.\ Suppl.\  {\bf 78}, 1 (1984).


\bibitem{MFB}
  V.~F.~Mukhanov, H.~A.~Feldman and R.~H.~Brandenberger,
  Phys.\ Rept.\  {\bf 215}, 203 (1992).

\bibitem{LLBook}
A.~R.~Liddle and D.~H.~Lyth,
\emph{Cosmological inflation and large-scale structure}, CUP,
Cambridge, UK (2000).




\bibitem{Mukhanov}
V.~F.~Mukhanov, L.~R.~W.~Abramo and R.~H.~Brandenberger,
Phys.\ Rev.\ Lett.\  {\bf 78}, 1624 (1997)
[arXiv:gr-qc/9609026].


\bibitem{Gauge1}
M.~Bruni, S.~Matarrese, S.~Mollerach and S.~Sonego,
Class.\ Quant.\ Grav.\  {\bf 14}, 2585 (1997) [arXiv:gr-qc/9609040].


\bibitem{Met2}
V.~Acquaviva, N.~Bartolo, S.~Matarrese and A.~Riotto,
Nucl.\ Phys.\ B {\bf 667}, 119 (2003) [arXiv:astro-ph/0209156].


\bibitem{Noh}
H.~Noh and J.~c.~Hwang,
Phys.\ Rev.\ D {\bf 69}, 104011 (2004).


\bibitem{MW2004}
K.~A.~Malik and D.~Wands,
Class.\ Quant.\ Grav.\  {\bf 21}, L65 (2004)
[arXiv:astro-ph/0307055].


\bibitem{Rigopoulos:2004gr}
  G.~I.~Rigopoulos and E.~P.~S.~Shellard,
  JCAP {\bf 0510}, 006 (2005)
  [arXiv:astro-ph/0405185].



\bibitem{Bartolo:2004if}
N.~Bartolo, E.~Komatsu, S.~Matarrese and A.~Riotto,
Phys.\ Rept.\  {\bf 402}, 103 (2004)
[arXiv:astro-ph/0406398].



\bibitem{Bart1}
  N.~Bartolo, S.~Matarrese and A.~Riotto,
  Phys.\ Rev.\ Lett.\  {\bf 93}, 231301 (2004)
  [arXiv:astro-ph/0407505].


\bibitem{Nak3}
  K.~Nakamura,
  Prog.\ Theor.\ Phys.\  {\bf 113} (2005) 481
  [arXiv:gr-qc/0410024].


\bibitem{LMS}
  D.~H.~Lyth, K.~A.~Malik and M.~Sasaki,
  JCAP {\bf 0505}, 004 (2005)
  [arXiv:astro-ph/0411220].


\bibitem{Tom1}
  K.~Tomita,
  Phys.\ Rev.\  D {\bf 71} (2005) 083504
  [arXiv:astro-ph/0501663].


\bibitem{Lyth:2005du}
D.~H.~Lyth and Y.~Rodriguez,
Phys.\ Rev.\  D {\bf 71}, 123508 (2005)
[arXiv:astro-ph/0502578].


\bibitem{Rigopoulos:2005xx}
  G.~I.~Rigopoulos, E.~P.~S.~Shellard and B.~J.~W.~van Tent,
  Phys.\ Rev.\  D {\bf 73}, 083521 (2006)
  [arXiv:astro-ph/0504508].


\bibitem{Malik2005}
K.~A.~Malik,
JCAP {\bf 0511}, 005 (2005) [arXiv:astro-ph/0506532v5].


\bibitem{Bart2}
  N.~Bartolo, S.~Matarrese and A.~Riotto,
  JCAP {\bf 0605}, 010 (2006)
  [arXiv:astro-ph/0512481].



\bibitem{Nak5}
  K.~Nakamura,
  Prog.\ Theor.\ Phys.\  {\bf 117} (2007) 17
  [arXiv:gr-qc/0605108].


\bibitem{Malik2006}
K.~A.~Malik,
JCAP {\bf 0703}, 004 (2007)
[arXiv:astro-ph/0610864v5].


\bibitem{Cov1}
  D.~Langlois and F.~Vernizzi,
  JCAP {\bf 0702}, 017 (2007)
  [arXiv:astro-ph/0610064].


\bibitem{Ananda1}
  K.~N.~Ananda, C.~Clarkson and D.~Wands,
  Phys.\ Rev.\  D {\bf 75}, 123518 (2007)
  [arXiv:gr-qc/0612013].


\bibitem{Mena1}
  F.~C.~Mena, D.~J.~Mulryne and R.~Tavakol,
  Class.\ Quant.\ Grav.\  {\bf 24} (2007) 2721
  [arXiv:gr-qc/0702064].


\bibitem{Ananda2}
  T.~C.~Lu, K.~Ananda and C.~Clarkson,
  arXiv:0709.1619 [astro-ph].






\bibitem{Maldacena}
J.~Maldacena,
JHEP {\bf 0305}, 013 (2003) [arXiv:astro-ph/0210603].


\bibitem{BartNG5}
  N.~Bartolo, E.~Komatsu, S.~Matarrese and A.~Riotto,
  Phys.\ Rept.\  {\bf 402}, 103 (2004)
  [arXiv:astro-ph/0406398].


\bibitem{Seery:2005wm}
D.~Seery and J.~E.~Lidsey,
JCAP {\bf 0506}, 003 (2005) [arXiv:astro-ph/0503692].


\bibitem{Seery:2005gb}
D.~Seery and J.~E.~Lidsey,
JCAP {\bf 0509}, 011 (2005) [arXiv:astro-ph/0506056].


\bibitem{Vernizzi:2006ve}
  F.~Vernizzi and D.~Wands,
  JCAP {\bf 0605}, 019 (2006)
  [arXiv:astro-ph/0603799].



\bibitem{Seery:2006vu}
  D.~Seery, J.~E.~Lidsey and M.~S.~Sloth,
  JCAP {\bf 0701}, 027 (2007)
  [arXiv:astro-ph/0610210].


\bibitem{Battefeld:2006sz}
  T.~Battefeld and R.~Easther,
  JCAP {\bf 0703}, 020 (2007)
  [arXiv:astro-ph/0610296].

\bibitem{Seery:2006js}
  D.~Seery and J.~E.~Lidsey,
  JCAP {\bf 0701}, 008 (2007)
  [arXiv:astro-ph/0611034].

\bibitem{Battefeld:2007en}
  D.~Battefeld and T.~Battefeld,
  JCAP {\bf 0705}, 012 (2007)
  [arXiv:hep-th/0703012].













\bibitem{Lidsey:2007gq}
  J.~E.~Lidsey and I.~Huston,
  JCAP {\bf 0707}, 002 (2007)
  [arXiv:0705.0240 [hep-th]].

\bibitem{Baumann:2006cd}
  D.~Baumann and L.~McAllister,
  Phys.\ Rev.\  D {\bf 75}, 123508 (2007)
  [arXiv:hep-th/0610285].

\bibitem{Peiris:2007gz}
  H.~V.~Peiris, D.~Baumann, B.~Friedman and A.~Cooray,
  Phys.\ Rev.\  D {\bf 76}, 103517 (2007)
  [arXiv:0706.1240 [astro-ph]].

\bibitem{Bean:2007eh}
  R.~Bean, X.~Chen, H.~V.~Peiris and J.~Xu,
  arXiv:0710.1812 [hep-th].





\bibitem{Lyth:2002my}
  D.~H.~Lyth, C.~Ungarelli and D.~Wands,
  Phys.\ Rev.\  D {\bf 67}, 023503 (2003)
  [arXiv:astro-ph/0208055].


\bibitem{Bartolo:2003jx}
  N.~Bartolo, S.~Matarrese and A.~Riotto,
  Phys.\ Rev.\  D {\bf 69}, 043503 (2004)
  [arXiv:hep-ph/0309033].


\bibitem{Malik:2006pm}
  K.~A.~Malik and D.~H.~Lyth,
  JCAP {\bf 0609}, 008 (2006)
  [arXiv:astro-ph/0604387].


\bibitem{Sasaki:2006kq}
  M.~Sasaki, J.~Valiviita and D.~Wands,
  Phys.\ Rev.\  D {\bf 74}, 103003 (2006)
  [arXiv:astro-ph/0607627].





\bibitem{Sasaki1986}
M.~Sasaki,
Prog.\ Theor.\ Phys.\  {\bf 76}, 1036 (1986).


\bibitem{Mukhanov88}
V.~F.~Mukhanov,
Sov.\ Phys.\ JETP {\bf 67}, 1297 (1988) [Zh.\ Eksp.\ Teor.\ Fiz.\
{\bf 94N7}, 1 (1988)].



\bibitem{Boundary}
J.~W.~York,
Phys.\ Rev.\ Lett.\ {\bf 28}, 1082 (1972);
G.~W.~Gibbons and S.~W.~Hawking,
Phys.\ Rev.\ D\ {\bf 15}, 2752 (1977).


\bibitem{Chamblin:1999ya}
  H.~A.~Chamblin and H.~S.~Reall,
  Nucl.\ Phys.\  B {\bf 562}, 133 (1999)
  [arXiv:hep-th/9903225].


\bibitem{Zaldarriaga}
M.~Zaldarriaga,
Phys.\ Rev.\  D {\bf 69}, 043508 (2004)
[arXiv:astro-ph/0306006].



\end{thebibliography}
\end{document}